\documentclass[usenatbib]{mnras}
\usepackage[british]{babel}             
\usepackage{newtxtext}                  
\usepackage[slantedGreek]{newtxmath}    
\usepackage{graphicx}                   
\def\HII{\ion{H}{II}}
\def\SNR(#1.#2)#3(#4.#5){{G#1${\cdot}$#2$#3$#4${\cdot}$#5}}

\def\cosec{\mathop{\mathrm{cosec}}}
\let\ga=\gtrsim  
\hypersetup{pdfauthor={D. A. Green and S. Roy},
  pdftitle={GMRT observations of the radio trail from CXOU J163802.6-471358},
  pdfkeywords={radio continuum: ISM; ISM: individual objects: CXOU
  J163802.6-471358, G338.1+0.4; pulsars: general; ISM: supernova remnants},
  bookmarksnumbered=true}
\def\CXOU{CXOU J163802.6$-$471358}
\volume{{\rm in press}}
\title{GMRT observations of the radio trail from CXOU J163802.6$-$471358}

\author[Green \& Roy]{D.~A.~Green$^1$\thanks{email: {\tt dag@mrao.cam.ac.uk},
   {\tt roy@ncra.tifr.res.in}} and S.~Roy$^2$\footnotemark[1]\\
$^1$Astrophysics Group, Cavendish Laboratory, 19 J.~J.~Thomson Avenue,
    Cambridge CB3 0HE\\
$^2$National Centre for Radio Astrophysics -- Tata Institute of
    Fundamental Research, Pune, Pune University Campus,\\
    Post Bag 3, Ganeshkhind Pune 411007, INDIA}

\date{Accepted ---; received ---; in original form ---}

\pagerange{\pageref{firstpage}--\pageref{lastpage}}

\pubyear{2023}
\begin{document}

\label{firstpage}

\maketitle

\begin{abstract}
The X-ray source {\CXOU} is thought to be a pulsar wind nebula (PWN), as
it shows an extended, $\approx 40$~arcsec trail from a compact source.
Here we present GMRT observations of this source at 330 and 1390~MHz,
which reveal a remarkable linear radio trail $\approx 90$~arcsec in
extent. Although the radio trail points back to the supernova
remnant (SNR) G338.1$+$0.4, $\approx 50$~arcmin from {\CXOU},
associating it with this remnant would require a very large velocity for
the pulsar. There are no known galactic SNRs close to the PWN and radio
trail. No pulsar has yet been identified in CXOU J163802.6$-$471358, but
if one could be found, this would allow more quantitative studies of the
PWN and radio trail to be made.
\end{abstract}


\section{Introduction}\label{s:introduction}

Supernovae (SNe) can produce both extended supernova remnants (SNRs) and
also pulsars in the case of core collapse `type II' SNe. Pulsars
dissipate most of their rotational energy through relativistic winds,
and in some circumstances these winds can produce a luminous nebula
around the pulsar which is known as a pulsar wind nebula (PWN, see
\citealt{2006ARA&A..44...17G} for a review). In some, younger cases the
PWN dominates the SNR, e.g.\ Crab nebula ($=$G184.6$-$5.8) and 3C58
($=$G130.7$+$3.1), see \citet{2014RPPh...77f6901B, 2013A&A...560A..18K}
and references therein. In other, old cases the pulsar/PWN is within the
shell of the remnant, e.g.\ MSH 15-5{\it 6} ($=$G326.3$-$1.8), see
\cite{2000ApJ...543..840D, 2017ApJ...851..128T}.

\begin{figure*}
\centerline{\includegraphics[width=7.0cm]{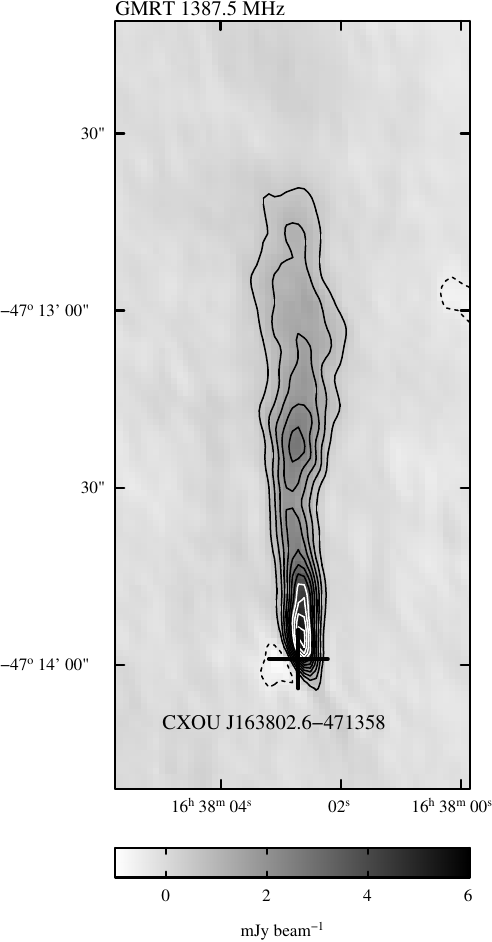}
     \qquad \includegraphics[width=7.0cm]{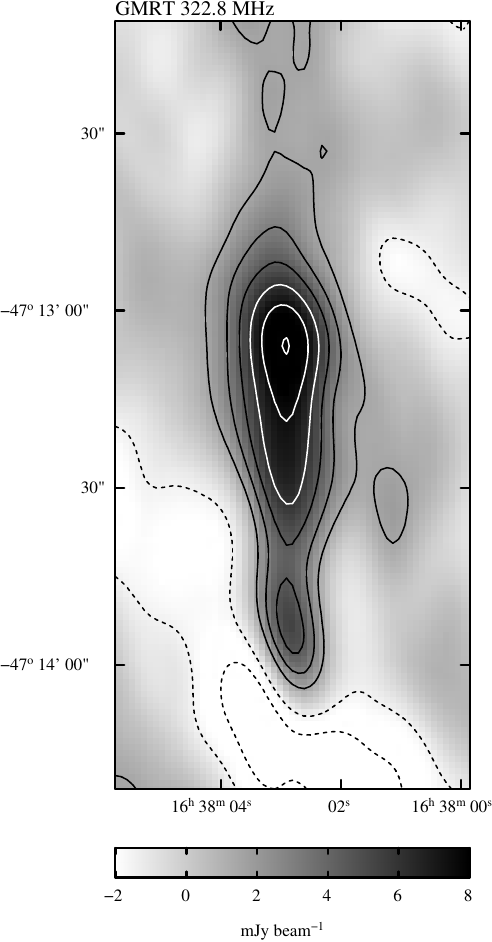}}
\caption{Image of radio trail from {\CXOU} at 1387.5 and 322.8~MHz from
GMRT observations, in J2000.0 equatorial coordinates. At 1387.5~MHz the
greyscale is $-1$ to $+6$ mJy~beam$^{-1}$ with contours every 0.5
mJy~beam$^{-1}$, and the resolution is $15.0 \times 6.1$~arcsec$^{2}$ at
a position angle of $8.9^\circ$ (E of N). At 322.8~MHz the greyscale is
$-2$ to $+8$ mJy~beam$^{-1}$ with contours every 1.5 mJy~beam$^{-1}$,
and the resolution $6.0 \times 3.1$~arcsec$^{2}$ at a position angle of
$24^\circ$. Negative contours are dashed.\label{f:GMRT}}
\end{figure*}

\begin{figure*}
\centerline{\includegraphics[width=11.0cm]{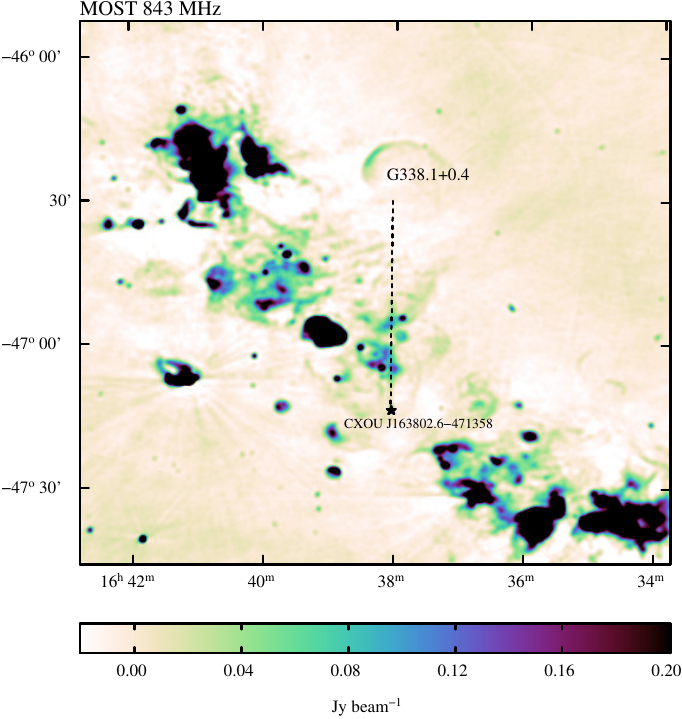}}
\caption{Molonglo Galactic Plane Survey \citep{2007MNRAS.382..382M}
observations, at 843~MHz of region surrounding {\CXOU}, in J2000.0
equatorial coordinates. The resolution is $45 \times 45 \cosec(|{\rm
Dec}|)$~arcsec$^2$, and the `cubehelix' colour scheme
\citep{2011BASI...39..289G} is from $-20$ to $200$ mJy~beam$^{-1}$. The
position of {\CXOU} is marked with a star, and the dotted line show the
alignment of the radio trail from it.\label{f:mgps2}}
\end{figure*}

\begin{figure}
\centerline{\includegraphics[width=8.5cm]{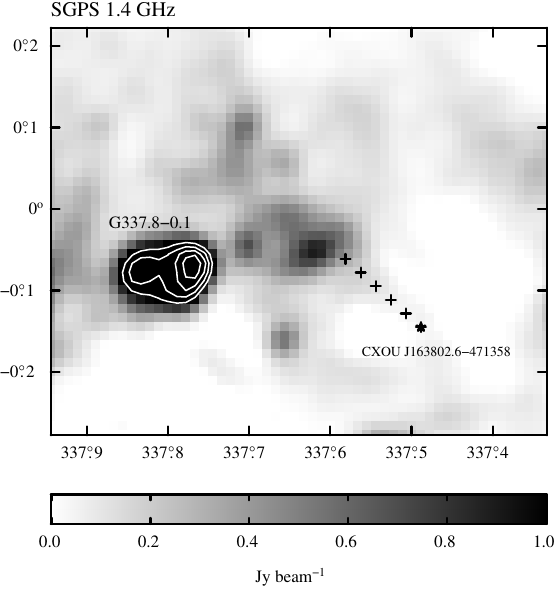}}
\medskip
\centerline{\includegraphics[width=8.5cm]{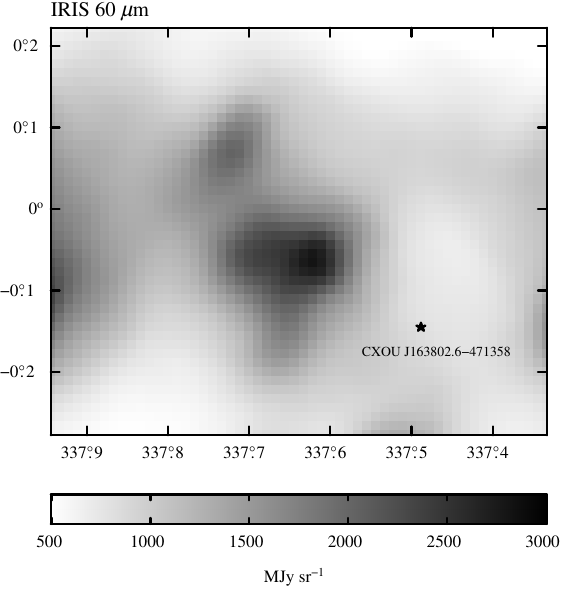}}
\caption{Radio and infra-red emission from the region to the north of
{\CXOU}, shown by a star, in Galactic coordinates. (top) SGPS 1.4-GHz
radio emission, with resolution $100$~arcsec
\citep{2006ApJS..167..230H}. The greyscale is 0.0 to 1.0 Jy~beam$^{-1}$,
with selected contours at 1.1, 1.4, 1.7 and 2.0 Jy~beam$^{-1}$ for the
bright emission from the SNR G337.8$-$0.1. The crosses mark the
extension of the observed radio trail from {\CXOU}. (bottom) IRIS
60-$\upmu$m emission, with resolution of $\approx 3$~arcmin
\citep{2005ApJS..157..302M}. The greyscale is 500 to 3000
MJy~sr$^{-1}$.\label{f:sgps-iris}}
\end{figure}

Some asymmetric SNe produce pulsars with large `kick' velocity (e.g.\
\citealt{2017ApJ...837...84J, 2017ApJ...844...84H}), and in cases where
a pulsar is moving through the surrounding medium supersonically, then
it forms a bow shock PWN due to ram pressure (e.g.\
\citealt{2006ARA&A..44...17G}). Such a PWN may be within the supernova
remnant shell, or if the kick velocity or age are large enough, outside
it. About sixty PWNs are known \citep{PWNcatalogue}, with only about 15
PWNs are seen to have the morphology of a bow shock PWN, e.g.\
G270.3$-$1.0, the `Lighthouse nebula' (=IGR J11014$-$6103) near the SNR
G290.1$-$0.8 ($=$MSH 11$-$6{\it 1}A), G315.9$-$0.0 ($=$ the `Frying
Pan'), G341.2$+$0.9, G359.2$-$0.8 ($=$ the `Mouse', e.g.\
\citealt{2009ApJ...706.1316H}), G5.27$-$0.9 ($=$ the `Duck'), and
G116.9$+$0.1 ($=$CTB~1, \citealt{2019ApJ...876L..17S,
2023ApJ...945..129K}). Of these, the `Frying Pan'
\citep{2009ApJ...703L..55C, 2012ApJ...746..105N} is particularly
striking, with a long radio trail extending $\approx 8$~arcmin from the
remnant shell, which has a radius of $\approx 6$~arcmin. The extent of
the radio trail is estimated to be about 20~pc, and the pulsar velocity
about 1000~km~s$^{-1}$ for a distance of 8~kpc (but also see
\citealt{2020A&A...639A..72W} who give a lower distance for
G315.9$-$0.0). Higher resolution radio images of this radio trail
\citep{2012ApJ...746..105N} show that it has a `kink', which reflects
variations in the medium the pulsar is moving through. In the case of
the `Lighthouse nebula', X-ray observations \citep{2016A&A...591A..91P}
show there is -- in addition to a PWN trail about 1~arcmin long -- there
is also a curved `jet' that is approximately perpendicular to the PWN
trail. The `jet' is fainter and longer than the PWN trail (at least
5~arcmin compared with PWN trail), and is well modelled by a helical
geometry.

The X-ray source {\CXOU} \citep{2014ApJ...796..105F,
2014ApJ...787..129J} is thought to be a PWN, as Chandra observations
show a faint, diffuse `tail' $\approx 40$~arcsec long extending to the
north of a compact source. {\CXOU} has Galactic coordinates of $l
\approx 337\fdg48$, $b \approx -0\fdg15$, so lies very close to the
Galactic plane. The number of Chandra counts in the `tail' is small,
only 110. \citeauthor{2014ApJ...787..129J} note also that there is a
suggestion from smoothed images that there may be a shorter, fainter
extension to west of the compact source -- i.e.\ perpendicular to the
main `tail' -- but the significance is of this is low ($2.9\sigma$),
given the small number of X-ray counts. Further,
\citeauthor{2014ApJ...787..129J} note this source is also detected by
XMM, and that there is a faint radio source in the Molonglo Galactic
Plane Survey (MGPS-2, \citealt{2007MNRAS.382..382M}) survey at 843-MHz.
The XMM source was subsequently catalogued as 3XMM J163802.6$-$471357 in
the 3rd XMM--Newton serendipitous survey \citep{2016A&A...590A...1R}.
\citeauthor{2014ApJ...787..129J} describe the MGPS-2 radio source as
being a unresolved, and offset from the compact X-ray source by about
40~arcsec. However, the resolution of the MGPS-2 image is rather low
($43 \times 64$~arcsec$^2$ elongated north--south). The MGPS-2 image
also shows a possible faint extension $\sim 3$~arcmin to the north,
although there appear to be artefacts in the image, which is in a
complex region of the Galactic plane. Subsequently {\CXOU} has also been
detected in hard X-rays by NuSTAR \citep{2017ApJS..229...33F}. To date
no pulsar has been detected within {\CXOU}.

In order to clarify the nature of the faint MGPS radio source, we have
obtained higher resolution radio observations of it, at both 322 and
1390-MHz, with Giant Meterwave Radio Telescope (GMRT) which reveal it to
be a remarkably long linear trail. Section~\ref{s:observations}
describes the observations and presents the resulting images. These are
discussed in Section~\ref{s:discussion}, and our conclusions presented
in Section~\ref{s:conclusions}.

\section{Observations and Results}\label{s:observations}

The GMRT -- see \citet{2002IAUS..199..439R} -- is a synthesis telescope
consisting of thirty 45-m antennas, that provides baselines up to $\sim
25$~km. {\CXOU} was observed with the GMRT at both 330 and 1390~MHz, in
separate observing sessions on 2015 July 9th and 23rd respectively. Each
observation was made with a bandwidth of 32~MHz, divided into 256
channels. Since  CXOU J163802.6$-$471358 is at a declination of $\approx
-47^\circ$ and the GMRT is at a latitude of $\approx 19^\circ$, these
observations were made at low elevations ($17^\circ$ to $24^\circ$),
with about 2.5 hours integration time on the source. Each observing
session included a short observation of 3C286, which was used for flux
density calibration, and interleaved observations of nearby secondary
calibrators. The data was analysed following the standard procedures in
AIPS (e.g.\ \citealt{2003ASSL..285..109G}).

Because {\CXOU} is at such a low declination, there was no known
standard 330-MHz secondary calibrator within 15 deg of it. However, a
compact, bright source J1627$-$3953 was identified from the NRAO VLA Sky
Survey (NVSS, \citealt{1998AJ....115.1693C}) at 1.4~GHz, which was found
to be compact and bright at 330~MHz with the GMRT, with a flux density
of about 6~Jy. This was used as the secondary calibrator for the 330-MHz
observations, with the model of the calibrator also including two much
weaker sources in the field of view. J1627$-$3953 was observed
approximately about every thirty minutes, and was also used as the
bandpass calibrator at 330~MHz. After calibration and editing of the
330-MHz data, 10 adjacent frequency channels were averaged providing a
channel width of 1.25 MHz in the output data. This reduces the data
volume significantly, while keeping bandwidth smearing smaller than the
synthesized beam during imaging up to half power point of primary beam
of the GMRT. Imaging of {\CXOU} at 330~MHz is not straightforward, due
to large field of view of the GMRT and the position of {\CXOU} in the
Galactic plane near various sources of emission on a wide range of
angular scales. Initial images were made, using multiple facets to cover
the field of view (see \citealt{KoganGreisen2009}), and
improved by two cycles of phase-only self-calibration. The final image
was made in two steps.
\begin{enumerate}
\item A high resolution CLEANed image of all the facets were made with a
short $uv$ cut-off of 1 k$\lambda$, to exclude extended emission of
angular size $\ga 3$~arcmin. All the significant CLEAN components
produced in all the facets were then subtracted from the self-calibrated
$uv$ data.
\item A low resolution image from the resulting $uv$-data was then made
with no $uv$ cutoff, and significant CLEAN components from this image
were then subtracted from $uv$ data used in first step.
Separating out the short spacings reduces the effect of the
extended background Galactic emission.
These data were then used
to image the {\CXOU} for baselines $> 0.8$~k$\lambda$.
\end{enumerate}

In this processing some end channels were removed, as the ends of the
bandpasses have poor responses, giving an effective bandwidth of 28~MHz,
centred at 322.8~MHz. The resulting image of {\CXOU} is shown in
Fig.~\ref{f:GMRT}. This has a resolution of $15.0 \times
6.1$~arcsec$^2$, with an r.m.s.\ noise of $\approx 1$~mJy~beam$^{-1}$.
As noted above, because {\CXOU} is in the Galactic plane, the local
baselevel of the resulting image is not well defined.

At 1390~MHz the compact source J1626$-$2951 was observed approximately
every 40~min, and was used as the secondary calibrator. The observations
were calibrated using standard techniques. The observations of 3C286
were used to derive antenna based bandpass corrections. Several central
channels were averaged together, and the observations of J1626$-$298
were used to calibrate the amplitude and phase of the antennas through
the observations. The derived antenna-based bandpass, and
amplitude/phase calibrations were then applied to the observations of
{\CXOU}.

For the 1390~MHz observations 240 channels were retained, as the ends of
the bandpasses have poor responses, giving an effective bandwidth of
30~MHz, centred at 1387.5~MHz. Adjacent groups of 10 channels were
averaged together for imaging. Several cycle of self-calibration were
applied, adjusting the phase only on timescales of 11, 5 and 1.5~min,
and then one cycles of phase and amplitude self-calibration was applied,
on a timescale of 1.5~min. The resulting image of {\CXOU} is shown in
Fig.~\ref{f:GMRT}. This has a resolution of $10.3 \times
5.3$~arcsec$^2$, with an r.m.s.\ noise of $\approx 0.1$~mJy~beam$^{-1}$.

\section{Discussion}\label{s:discussion}

These GMRT observations reveal a remarkably linear radio trail that
extents $\approx 90$~arcsec directly to the north of the X-ray compact
source position, see Fig.~\ref{f:GMRT}. These results clearly
show radio emission from the position of the compact X-ray
source CXOU J163802.6$-$471358, with no offset (the `offset' noted by
\citeauthor{2014ApJ...787..129J} being due to the low resolution of the
MGPS-2 radio image). The radio trail broadens and fades away from the
position of {\CXOU}. As noted above, there is a suggestion of a fainter
X-ray trail perpendicular to the main trail from the compact X-ray
source (cf.\ the `Lighthouse nebula'), but there is no indication of any
radio emission in a perpendicular direction.
Determining the spectral index of the source is limited due to
variations in the background emission near source in the lower frequency
image. Its spectral index is consistent with being flat, but
with a large uncertainty (about 0.4) in the spectral index.

Assuming that {\CXOU} contains a pulsar, and the radio trail is due to
ram pressure stripping of the pulsar wind nebula moving supersonically
through the ISM, then it would be expected that this pulsar originated
in a SNR directly to the north of the radio trail. There are no
catalogued Galactic SNRs nearby, i.e.\ within a few tens of arcmin, to
the north of the radio trail. The closest known SNR is G337.8$-$0.1 near
$16^{\rm h}39^{\rm m}$, $-47^\circ$ (e.g.\ \citealt{1970AuJPA..14..133S,
1996A&AS..118..329W}), which is thought most likely to be in the Norma
II arm, at a distance of about 11~kpc \citep{2007A&A...468..993K}.
However, this is not aligned with the direction of the radio trail, but
is misaligned by about $30$~deg.

There \emph{is} a catalogued remnant -- G338.1$+$0.4, near $16^{\rm
h}38^{\rm m}$, $-46^\circ25'$ (e.g.\ \citealt{1970AuJPA..14..133S,
1979A&AS...38...39Z, 1996A&AS..118..329W}) -- that does lies due north
of {\CXOU}, i.e.\ along the line traced by the radio trail, but this is
quite far away. G338.1$+$0.4 is a slightly elongated shell remnant,
$\approx 16 \times 14$~arcmin$^2$ in extent at radio wavelengths (e.g.\
\citealt{1996A&AS..118..329W}), and has been detected optically
\citep{1979A&AS...38...39Z}. Fig.~\ref{f:mgps2} shows radio
image of the region of the Galactic plane from the Molonglo Galactic
Plane Survey (MGPS-2, see \citealt{2007MNRAS.382..382M}) near CXOU
J163802.6$-$471358, including G338.1$+$0.4. The centroid of G338.1$+$0.4
is $l \approx 338\fdg05$, $b \approx 0\fdg45$, which is about $\sim
0\fdg8$ from {\CXOU}, i.e.\ about 7 times the radius of G338.1$+$0.4.
Hence, if {\CXOU} was born by the supernova that produced G338.1$+$0.4,
then the velocity of pulsar in the plane of the sky would have to be 7
times the average expansion velocity of the remnant. No direct distance
measurement is available for G338.1$+$0.4, but the fact that it is
visible optically suggests that it is not very distant. If G338.1$+$0.4
is at 5~kpc, would have a radius of $\approx 12$~pc, and for a nominal
SN energy of $10^{44}$~J and an ISM density of 1~cm$^{-3}$, a Sedov
evolution model for the remnant would imply an age of about 8000 yr and
an average expansion velocity of about 1400~km~s$^{-1}$. This would
require an extremely large pulsar transverse (i.e.\ in the plane of the
sky) velocity of 9800~km~s$^{-1}$. This is much larger than the highest
pulsar velocities known, e.g.\ \citet{2005ApJ...630L..61C}, who measure
a transverse velocity of $1083^{+103}_{-90}$ km~s$^{-1}$ for the pulsar
B1508$+$55 from parallax and proper motion observations. In the case of
the `Lighthouse nebula', higher velocities of 2400--2900 km~s$^{-1}$
have been suggested for IGR J11014$-$6103 \citep{2012ApJ...750L..39T},
depending on the distance and age of the remnant G290.1$-$0.8. More
recently \citep{2016A&A...591A..91P} use a somewhat a smaller distance
to the remnant, but still derive a large pulsar velocity of $\approx
1000$ km~s$^{-1}$). Also, a velocity as large as 9800~km~s$^{-1}$ is not
possible to reconcile with a standard SN explosion, since the kinetic
energy of a 1.5~$M_\odot$ neutron star moving at 9800~km~s$^{-1}$ is
$\approx 1.4 \times 10^{44}$~J, which is comparable with the nominal
total SN energy.

Thus, the pulsar and {\CXOU} must have been born in a different SNR than
G338.1$+$0.4. The MGPS-2 image in Fig.~\ref{f:mgps2} show a complex
region of radio features near $16^{\rm h}38^{\rm m}15^{\rm s}$,
$+47^\circ3'$ (or $l \approx 337\fdg65$, $b \approx-0\fdg05$) which lie
to the north of CXOU J163802.6$-$471358, much closer than G338.1$+$0.4
to the radio trail. However, there are no catalogued SNRs in this region
\citep{2019JApA...40...36G}, nor are there any reported SNR candidates
(see Section 2.3 of \citealt{Green22}).
Fig.~\ref{f:sgps-iris} shows SGPS (see
\citealt{2006ApJS..167..230H}) radio and IRIS (see
\citealt{2005ApJS..157..302M}) 60~$\upmu$m infra-red images of this
region. These show that the radio emitting features, apart from
G337.8$-$0.1, are all associated with infra-red emission. Indeed in the
WISE {\HII} region catalogue \citep{2014ApJS..212....1A} includes
several known, candidate and `radio quiet' {\HII} regions near
$l=337\fdg65$, $b=-0\fdg05$. These include G337.665$-$0.048 listed in
\citep{1987A&A...171..261C}, and an ultracompact {\HII} region
G337.6156$-$0.0606 identified from IRAS observations
\citep{1997MNRAS.291..261W, 1998MNRAS.301..640W}.
Thus it is likely that the parent SNR of {\CXOU} is so old, and
hence faint at radio wavelengths that it is not identifiable. For an
age of $10^{5}$~yr, and pulsar speed of $\sim 200$~km~s$^{-1}$
in the plane of the sky (e.g.\ \citealt{2020MNRAS.494.3663I}), this
corresponds to a linear distance of 21~pc (or 14~arcmin at 5~kpc).

As yet no pulsar has yet been directly identified in
{\CXOU}, which means that no direct proper motion is
available, nor is a distance estimate from the pulsar dispersion
measure. Identification of the presumed pulsar
in {\CXOU} is needed for better understanding of any associated
parent SNR.

\section{Conclusions}\label{s:conclusions}

GMRT observations of the X-ray source {\CXOU}, which is thought to be a
PWN, at 330 and 1390~MHz, reveal a remarkable linear radio trail
extending $\approx 90$~arcsec to the north. Such a PWN trail is expected
to point back to the supernova remnant that produced the pulsar. The
closest known SNR to {\CXOU} is G337.8$-$0.1, but this is misaligned
with the radio trail by about 30~deg. There is a known SNR,
G338.1$+$0.4, which is aligned with the radio trail. But it is $\approx
50$~arcmin from {\CXOU}, which would require a very high pulsar
transverse velocity for it to be where the pulsar in {\CXOU} to have
been born. Thus, the parent SNR of this PWN is not currently identified.
Clearly, an identification of a pulsar would allow more quantitative
studies of the PWN and radio trail to be made.

\section*{Acknowledgements}

We thank the staff of the GMRT who have made these observations
possible. The GMRT is run by the National Centre for Radio Astrophysics
of the Tata Institute of Fundamental Research. We also thank the NCRA
Director for the allocation of discretionary time on the GMRT, and DAG
thanks NCRA for their hospitality during a sabbatical visit, when this
study was initiated.

\section*{Data Availability}

The GMRT data underlying this article are available from the GMRT Online
Archive (\url{https://naps.ncra.tifr.res.in/goa/}) for ProposalID
ddtB173. The other data use in Figs.~\ref{f:mgps2} and \ref{f:sgps-iris}
are available from
\url{http://www.astrop.physics.usyd.edu.au/mosaics/Galactic/},
\url{https://www.atnf.csiro.au/research/HI/sgps/fits_files.html} and
\url{https://irsa.ipac.caltech.edu/data/IRIS/images/}).

%
%

\bsp

\label{lastpage}
\end{document}